\documentclass[12pt,journal,draftcls,a4paper,onecolumn]{IEEEtran}

\pdfoutput=1

\usepackage{algorithm}
\usepackage{algorithmic}
\usepackage{graphicx}
\usepackage[cmex10]{amsmath}
\usepackage{amsfonts}
\usepackage{amssymb}
\usepackage{subfigure}
\usepackage{color}
\usepackage[all,poly]{xy}
\usepackage{multirow}
\usepackage{algorithm}
\usepackage{algorithmic}
\usepackage[noadjust]{cite}
\usepackage{verbatim}
\usepackage{dsfont}
\usepackage{breqn}
\usepackage{epic,eepic,pstricks}
\usepackage{dsfont}
\usepackage{subeqnarray}
\usepackage{cleveref}
\usepackage{lipsum}
\usepackage{relsize}
\usepackage{graphicx}

\graphicspath{{figures/}}

\newcommand{\bx}{\boldsymbol{x}}

\newcommand{\bv}{\boldsymbol{v}}
\newcommand{\by}{\boldsymbol{y}}
\newcommand{\bz}{\boldsymbol{z}}

\newcommand{\bmu}{\boldsymbol{\mu}}
\newcommand{\argmax}{\operatornamewithlimits{argmax}}
\newcommand{\argmin}{\operatornamewithlimits{argmin}}

\title{Fast unsupervised Bayesian image segmentation with adaptive spatial regularisation}
\author{Marcelo Pereyra and Steve McLaughlin

\thanks{This work was funded in part by the SuSTaIN program - EPSRC grant EP/D063485/1 - at the Department of Mathematics, University of Bristol, in part by a postdoctoral fellowship from French Ministry of Defence, and in part by the EPSRC via grant EP/J015180/1.This paper was presented in part at EUSIPCO'14 in Lisbon September 2014.}
\thanks{Marcelo Pereyra holds a Marie Curie Intra-European Fellowship for Career Development at the University of Bristol, School of Mathematics, University Walk, BS8 1TW, UK (e-mail:
marcelo.pereyra@bristol.ac.uk).}
\thanks{Steve McLaughlin is with Heriot Watt University, Engineering and Physical Sciences, Edinburgh, EH14 4AS, UK (e-mail: s.mclaughlin@hw.ac.uk).}}

\begin{document}

\maketitle

\begin{abstract}
This paper presents a new Bayesian estimation technique for hidden Potts-Markov random fields with unknown regularisation parameters, with application to fast unsupervised $K$-class image segmentation. The technique is derived by first removing the regularisation parameter from the Bayesian model by marginalisation, followed by a small-variance-asymptotic (SVA) analysis in which the spatial regularisation and the integer-constrained terms of the Potts model are decoupled. The evaluation of this SVA Bayesian estimator is then relaxed into a problem that can be computed efficiently by iteratively solving a convex total-variation denoising problem and a least-squares clustering (K-means) problem, both of which can be solved straightforwardly, even in high-dimensions, and with parallel computing techniques. This leads to a fast fully unsupervised Bayesian image segmentation methodology in which the strength of the spatial regularisation is adapted automatically to the observed image during the inference procedure, and  that can be easily applied in large $2$D and $3$D scenarios or in applications requiring low computing times. Experimental results on real images, as well as extensive comparisons with state-of-the-art algorithms, confirm that the proposed methodology offer extremely fast convergence and produces accurate segmentation results, with the important additional advantage of self-adjusting regularisation parameters.
\end{abstract}


\section{Introduction}
Image segmentation is a canonical inverse problem which involves classifying image pixels into clusters that are spatially coherent and have well defined boundaries. It is widely accepted that this task can be formulated as a statistical inference problem and most state-of-the-art image segmentation methods compute solutions by performing statistical inference (e.g., computing penalized maximum likelihood or maximum-a-posteriori estimates). In this paper we focus on new Bayesian computation methodology for hidden Potts-Markov random fields (MRFs) \cite{LiBook}, a powerful class of statistical models that is widely used in Bayesian image segmentation methods (see \cite{Eches2010tgrs,Ayasso2010,PereyraTMIC2011,Vincent2010}  for recent examples in hyperspectral, non destructive testing, ultrasound, and fMRI imaging). 

Despite the wide range of applications, performing inference on hidden Potts MRFs remains a computationally challenging problem. In particular, computing the maximum-a-posteriori (MAP) estimator for these models is generally NP-hard, and thus most image processing methods compute approximate estimators. This has driven the development of efficient approximate inference algorithms, particularly over the last decade. The current predominant approaches for approximate inference on MRFs are based on convex models and convex approximations that can be solved efficiently by convex optimisation \cite{Kolmogorov:09, Komodakis:2011, bioucas:2014}, and on approximate estimators computed with graph-cut \cite{Boykov:2001,Kolmogorov:04} and message passing algorithms \cite{Kolmogorov:2006, Felzenszwalb:2006, Szeliski:2008}. In a similar fashion, modern algorithms to solve active contour models, the other main class of models for image segmentation, are also principally based on convex relaxations and convex optimisation \cite{Bresson2007,Raymond2013} and on Riemannian steepest descent optimisation schemes \cite{PereyraSTSP2013,pereyra:2015,Bar2009,Sundaramoorthi2009}.

An important limitation of these computationally efficient approaches is that they are supervised, in the sense that require practitioners to specify the value of the regularisation parameter of the Potts MRF.  However, it is well known that appropriate values for regularisation parameters can be highly image dependent and sometimes difficult to select a priori, thus requiring practitioners to set parameter values heuristically or by visual cross-validation. The Bayesian framework offers a range of strategies to circumvent this problem and to design unsupervised image segmentation inference procedures that self-adjust their regularisation parameters. Unfortunately, the computations involved in these inferences are beyond the scope of existing fast approximate inference algorithms. As a consequence, unsupervised image segmentation methods have to use more computationally intensive strategies such as Monte Carlo approximations \cite{Pereyra_TIP_2013, Pereyra_SSP_2014}, variational Bayes approximations  \cite{McGrory2009}, and EM algorithms based on mean-field like approximations \cite{Celeux2003, Forbes2007}.

In this paper we propose a highly efficient Bayesian computation approach specifically designed for performing approximate inference on hidden Potts-Markov random fields with unknown regularisation parameters, with application to fast unsupervised $K$-class image segmentation. A main originality of our development is to use a small-variance-asymptotic (SVA) analysis to design an approximate MAP estimator in which the spatial regularisation and the integer-constrained terms of the Potts model are decoupled. The evaluation of this SVA Bayesian estimator can then be relaxed into a problem that can be computed efficiently by iteratively solving a convex total-variation denoising problem and a least-squares clustering (K-means) problem, both of which can be solved straightforwardly, even in high-dimensions, and with parallel computing techniques.

Small-variance asymptotics estimators were introduced in \cite{Broderick2013} as a computationally efficient framework for performing inference in Dirichlet process mixture models and have been recently applied to other important machine learning classification models such as the Beta process and sequential hidden Markov models \cite{NIPS2013_4913}, as well as to the problem of configuration alignment and matching \cite{mardiafest}. Here we exploit these same techniques for the hidden Potts MRF to develop an accurate and computationally efficient image segmentation methodology for the fully unsupervised case of unknown class statistical parameters (e.g., class means) and unknown Potts regularisation parameter. 

The paper is organised as follows: in Section II we present a brief background to Bayesian image segmentation using the Potts MRF. This then followed by a detailed development of our proposed methodology. In Section IV the methodology is applied to some standard example images and compared to other image segmentation approaches from the state of the art. Finally some brief conclusions are drawn in Section V. 

\section{Background}\label{Sec2}
We begin by recalling the standard Bayesian model used in image segmentation problems, which is based on a finite mixture model and a hidden Potts-Markov random field with known regularisation parameter $\beta$. For simplicity we focus on univariate Gaussian mixture models. However, the results presented hereafter can be generalised to all exponential-family mixture models (e.g., mixtures of multivariate Gaussian, Rayleigh, Poisson, Gamma, Binomial, etc.) by following the approach described in \cite{KeNIPS2012}.

Let $y_n \in \mathbb{R}$ denote the $n$th observation (i.e. pixel or voxel) in a lexicographical vectorized image $\boldsymbol{y}=(y_1,\ldots,y_N)^T \in \mathbb{R}^{N}$. We assume that $\boldsymbol{y}$ is made up by $K$ regions $\{\mathcal{C}_1, \ldots, \mathcal{C}_K\}$ such that the observations in the $k$th class are distributed according to the following conditional marginal observation model
\begin{eqnarray} \label{mixture}
y_n | n \in \mathcal{C}_k \sim \mathcal{N}(\mu_k,\sigma^2),
\end{eqnarray}
where $\mu_k \in \mathbb{R}$ represents the mean intensity of class $\mathcal{C}_k$. For identifiability we assume that $\mu_k \neq \mu_j$ for all $k \neq j$.

\begin{figure}
  \centerline{\includegraphics[width=7.5cm]{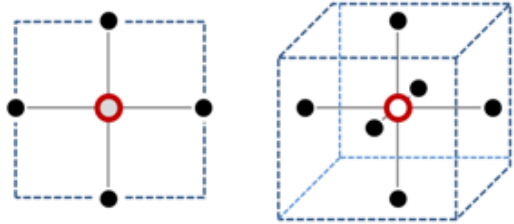}}
  \caption{4-pixel (left) and 6-voxel (right) neighborhood structures. The pixel/voxels considered appears as a void red circle whereas its neighbors are depicted in full black and blue.}\label{fig:neighborhood}
\end{figure}

To perform segmentation, a label vector $\boldsymbol{z}=\left(z_1,\ldots,z_N\right)^T$ is introduced to map or classify observations $\boldsymbol{y}$ to classes $\mathcal{C}_1, \ldots, \mathcal{C}_K$ (i.e., $z_n = k$ if and only if $n \in \mathcal{C}_k$). Assuming that observations are conditionally independent given $\bz$ and given the parameter vector $\bmu = (\mu_1,\ldots,\mu_K)$, the likelihood of $\by$ can be expressed as follows
\begin{eqnarray} \label{likelihood}
f(\by|\bz,\bmu) = \prod_{k=1}^K \,\,\prod_{n \in \mathcal{S}_k} p_{\mathcal{N}}(y_n|\mu_k,\sigma^2),
\end{eqnarray}
with $\mathcal{S}_k = \{n : z_n = k\}$.  A Bayesian model for image segmentation is then defined by specifying the prior distribution of the unknown parameter vector $(\bz,\bmu)$. The prior for $\boldsymbol{z}$ is the homogenous $K$-state Potts MRF \cite{Wu1982}
\begin{eqnarray} \label{Potts}
f(\boldsymbol{z}|\beta) = \frac{1}{C(\beta)} \exp{\left[\beta H(\boldsymbol{z})\right]},
\end{eqnarray}
with regularisation hyper-parameter $\beta \in \mathbb{R}^+$, Hamiltonian
\begin{eqnarray} \label{Hamiltonian}
    H(\boldsymbol{z}) = \sum_{n=1}^N\sum_{n'\in \mathcal{V}(n)} \delta(z_{n} == z_{n'}),
\end{eqnarray}
where $\delta(\cdot)$ is the Kronecker function and $\mathcal{V}(n)$ is the index set of the neighbors of the $n$th voxel (most methods use the $1$st order neighbourhoods depicted in Fig. \ref{fig:neighborhood}), and normalising constant (or partition function)
\begin{eqnarray} \label{Partition}
    C(\beta) = \sum_{\bz} \exp{\left[\beta H(\boldsymbol{z})\right]}.
\end{eqnarray}
Notice that the Potts prior \eqref{Potts} is defined conditionally to a given value of $\beta$. Most image segmentation methods based on this prior are supervised; i.e., assume that the value of $\beta$ is known and specified a priori by the practitioner. Alternatively, unsupervised methods consider that $\beta$ is unknown and seek to adjust its value automatically during the image segmentation procedure (this point is explained in detail in Section \ref{Sec3a}).

In a similar fashion, the class means are considered prior independent and assigned Gaussian priors $\mu_k \sim \mathcal{N}(0,\rho^2)$ with fixed variance $\rho^2$,
\begin{eqnarray} \label{fmu}
f(\bmu) = \prod_{k=1}^K p_{\mathcal{N}}(\mu_k|0,\rho^2).
\end{eqnarray}
(to simplify notation the dependence of distributions on the fixed quantity $\rho^2$ is omitted).

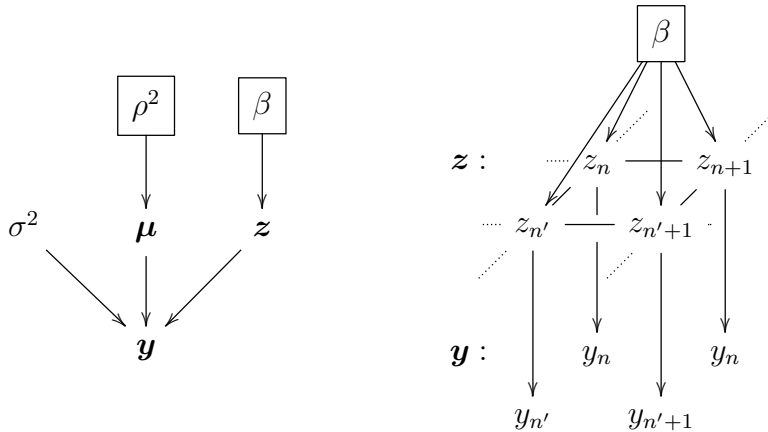
\begin{figure}[h!]
\begin{minipage}[a1]{.48\linewidth}
\centerline{ \xymatrix{
& *+[F]{\rho^2}\ar@{->}[d] & *+[F]{\beta}\ar@{->}[d] \\
\sigma^2 & \bmu\ar[d] & \bz \ar[dl]\\
& \by\ar@{<-}[ul] \\
}
}
\end{minipage}
\begin{minipage}[a1]{.48\linewidth}
\xymatrix@!0{
& & & *+[F]{\beta} \ar[ddr]\ar[ddl] \ar[ddd]\ar[dddll]& \\
& & & & &\\
\bz: & & z_n \ar@{.}[l] \ar@{.}[ur]\ar@{-}[rr]\ar@{->}'[d][ddd]
& & z_{n+1} \ar@{->}[ddd]\ar@{.}[r] \ar@{.}[ur]& & 
\\
& z_{n^\prime}  \ar@{.}[dl]\ar@{.}[l] \ar@{-}[ur]\ar@{-}[rr]\ar@{->}[ddd]
& & z_{n^\prime+1}  \ar@{-}[ur]\ar@{->}[ddd]\ar@{.}[r]\ar@{.}[dl]& 
\\
& & & & & 
\\
\by: & &  y_n 
& & y_n
\\
& y_{n^\prime} 
& & y_{n^\prime+1}
}
\end{minipage}
 \caption{[Left:] Directed acyclic graph of the standard Bayesian model for image segmentation (parameters with fixed values are represented using black boxes). [Right] Local hierarchical representation of the hidden Potts MRF and the observed image for $4$ neighbouring pixels.} \label{fig:DAG1}
\end{figure}

Then, using Bayes theorem and taking into account the conditional independence structure of the model (see Fig. \ref{fig:DAG1}), the joint posterior distribution of $(\boldsymbol{z},\mu)$ given $\by$ and $\beta$ can be expressed as follows
\begin{eqnarray} \label{posterior}
f\left(\bz,\bmu |\boldsymbol{y},\beta\right) \propto
f(\boldsymbol{y}|\boldsymbol{z},\bmu)f(\boldsymbol{z}|\beta)f(\bmu),
\end{eqnarray}
where $\propto$ denotes proportionality up to a normalising constant that can be retrieved by setting $\int f\left(\bz,\bmu |\boldsymbol{y},\beta\right) \textrm{d}\bz \textrm{d}\bmu = 1$. The graphical structure of this Bayesian model is summarised in Fig. \ref{fig:DAG1} below. Notice the Markovian structure of $\bz$ and that observations $y_n$ are conditionally independent given the model parameters $\bz$, $\bmu$ and $\sigma^2$.

Finally, given the Bayesian model \eqref{posterior}, a segmentation of $\by$ is typically obtained by computing the MAP estimator
\begin{eqnarray} \label{MAP}
\hat{\bz}_1,\hat{\bmu}_1 = \argmax_{\bz,\bmu}\, f\left(\bz,\bmu |\boldsymbol{y},\beta \right),
\end{eqnarray}
which can also be obtained by solving the equivalent optimisation problem
\begin{eqnarray} \label{MAP2}
\hat{\bz}_1,\hat{\bmu}_1 = \argmin_{\bz,\bmu}\, -\log f\left(\bz,\bmu |\boldsymbol{y}, \beta \right).
\end{eqnarray}
Unfortunately these optimisation problems are known to be NP-hard due to the combinatorial nature of the Potts Hamiltonian $H(\bz)$ defined in \eqref{Hamiltonian}. As mentioned previously, modern image segmentation methods based on \eqref{posterior} typically address this issue by using approximate (local) integer optimisation algorithms (e.g., graph-cut, message passing) \cite{Kolmogorov:04, Kolmogorov:2006, Felzenszwalb:2006}, and more recently with convex relaxations of the Potts model (see for instance \cite{Kolmogorov:09, Komodakis:2011}).

\section{Proposed method}\label{Sec3}
This section presents a highly computationally efficient approach for performing approximate inference on $\bz$ when the value of the regularisation parameter $\beta$ is unknown. The approach is based on a small-variance asymptotics (SVA) analysis combined with a convex relaxation and a pseudo-likelihood approximation of the Potts MRF. Our development has three main steps. In the first step we adopt a hierarchical Bayesian approach to remove $\beta$ from the model by marginalisation; because marginalising w.r.t. $\beta$ requires knowledge of the intractable Potts partition function \eqref{Partition} we use a pseudo-likelihood approximation. However, performing inference with the resulting marginalised model is still NP-hard. In the second part of our development we address this difficulty by using auxiliary variables and an SVA analysis to decouple the spatial regularisation and the integer-constrained terms of the Potts model. The evaluation of the resulting SVA Bayesian estimator is then relaxed into a problem that can be computed efficiently by iteratively solving a convex total-variation denoising problem and a least-squares clustering problem, both of which can be solved straightforwardly, even in high-dimensions, with parallel implementations of Chambolle's optimisation algorithm \cite{chambolle:2004} and of K-means \cite{macqueen:1967}.

\subsection{Marginalisation of the regularisation parameter $\beta$}\label{unsupervised}
Following a hierarchical Bayesian approach, we address the fact that the value of $\beta$ is unknown by modelling it as an additional random variable of the Bayesian model. Precisely, we assign $\beta$ a prior distribution $f(\beta)$ and define an augmented model that includes $\beta$ within its unknown parameter vector. By using Bayes' theorem we obtain the joint posterior distribution
\begin{eqnarray} \label{posterior2}
f\left(\bx,\bz,\bmu,\beta |\boldsymbol{y}\right) \propto f(\boldsymbol{y}|\bx)f(\bx|\boldsymbol{z},\bmu)f(\bmu)f(\boldsymbol{z}|\beta)f(\beta)
\end{eqnarray}
which includes $\beta$ as an unknown variable. The rationale for replacing the fixed regularisation parameter $\beta$ of \eqref{posterior} by a random variable with prior $f(\beta)$ is that it is often possible to specify this prior distribution such that the amount of regularisation enforced by the Potts MRF is driven by data and the impact of $f(\beta)$ on the inferences is minimal. At the same time, experienced practitioners with knowledge of good values of $\beta$ can specify $f(\beta)$ to exploit their prior beliefs. In this paper we use a gamma (hyper-)prior distribution
$$
f(\beta) =  \gamma^\alpha \beta^{\alpha-1} \exp{(-\gamma\beta)}\boldsymbol{1}_{\mathbb{R}^+}(\beta)/\Gamma(\alpha)
$$
because it has favourable analytical tractability properties that will be useful for our development (appropriate values for the fixed parameters $\alpha$ and $\gamma$ will be derived later through a small-variance asymptotics analysis).

Moreover, in order to marginalise $\beta$ from the model we notice that $\beta$ is conditionally independent of $\boldsymbol{y}$ given $\boldsymbol{z}$; to be precise, that $f\left(\bx,\bz,\bmu,\beta |\boldsymbol{y}\right) = f\left(\beta | \bz)f(\bx,\bz,\bmu |\boldsymbol{y}\right)$. Therefore, integrating $f\left(\bx,\bz,\bmu,\beta |\boldsymbol{y}\right)$ with respect to $\beta$ is equivalent to redefining the posterior distribution \eqref{posterior2} with the marginal prior $f(\bz) = \int_{\mathbb{R}^+} f(\bz,\beta) \textrm{d}\beta$. Evaluating this marginal prior exactly is not possible because it requires computing the normalising constant of the Potts model $C(\beta)$ defined in \eqref{Partition}, which is a reputedly intractable problem \cite{Pereyra_TIP_2013}. To obtain an analytically tractable approximation for this marginal prior we adopt a pseudo-likelihood approach \cite{besag:1986} and use the approximation $C(\beta) \propto \beta^{-N}$, leading to 
\begin{eqnarray}
\begin{split}
f(\bz)  & = \int_{\mathbb{R}^+} f(\bz,\beta) \textrm{d}\beta\\
	  & \propto \int_{\mathbb{R}^+}  \beta^N \exp{(\beta H(\boldsymbol{z}))} \beta^{\alpha-1} \exp{(-\gamma\beta)} \textrm{d}\beta\\
	  & \propto [\gamma-H(\boldsymbol{z})]^{-(\alpha + N)},
\end{split}
\end{eqnarray}
and to the following (marginal) posterior distribution
\begin{eqnarray} \label{posterior2}
\begin{split}
f\left(\bx,\bz,\bmu |\boldsymbol{y}\right) 	&\propto
								f(\bmu) \left(\gamma-H(\boldsymbol{z})\right)^{-(\alpha + N)} \prod_{k=1}^K \,\,\prod_{n \in \mathcal{S}_k} p_{\mathcal{N}}\left(y_n|\mu_k,\sigma^2\right),
\end{split}
\end{eqnarray}
that does not depend on the regularisation parameter $\beta$.

\subsection{Small-variance approximation}
The next step of our development is to conduct a small-variance asymptotics analysis on \eqref{posterior4} and derive the asymptotic MAP estimator of $\bx,\bz,\bmu$. We begin by introducing a carefully selected auxiliary vector $\bx$ such that $\by$ and $(\bz,\bmu)$ are conditionally independent given $\bx$, and that the posterior $f\left(\bx,\bz,\bmu |\boldsymbol{y}\right)$ has the same maximisers as \eqref{posterior} (after projection on the space of $(\bz,\bmu)$). More precisely, we define a random vector $\bx \in \mathbb{R}^{N}$ with degenerate prior
\begin{eqnarray}\label{priorX}
f(\bx|\boldsymbol{z},\bmu) = \prod_{k=1}^K \,\,\prod_{n \in \mathcal{S}_k} \delta(x_n - \mu_k),
\end{eqnarray} and express the likelihood of $\by$ given $\bx,\bz$ and $\bmu$ as 
$$
f(\by|\bx,\bz,\bmu) = f(\by|\bx) = \prod_{n=1}^N p_{\mathcal{N}}(y_n|x_n,\sigma^2).
$$
The prior distributions for $\bz$ and $\bmu$ remain as defined above. The joint posterior distribution of $\bx,\bz,\bmu$ is given by
\begin{eqnarray} \label{posterior3}
\begin{split}
f\left(\bx,\bz,\bmu, \beta |\boldsymbol{y} \right) 	&\propto
								f(\boldsymbol{y}|\bx)f(\bx|\boldsymbol{z},\bmu)f(\boldsymbol{z}|\beta)f(\bmu)\\
								&\propto
								\left[\prod_{k=1}^K \,\,\prod_{n \in \mathcal{S}_k} p_{\mathcal{N}}(y_n|x_n,\sigma^2)\delta(x_n - \mu_k)\right] f(\bmu) \left[\gamma-H(\boldsymbol{z})\right]^{-(\alpha + N)}.
\end{split}
\end{eqnarray}
Notice that from an inferential viewpoint \eqref{posterior3} is equivalent to \eqref{posterior2}, in the sense that marginalising $\bx$ in \eqref{posterior3} results in \eqref{posterior2}.

Moreover, we define $H^*(\boldsymbol{z})$ as the ``complement'' of the Hamiltonian $H(\boldsymbol{z})$ in the sense that for any $\bz \in [1,\ldots,K]^N$
$$
H(\boldsymbol{z}) + H^*(\boldsymbol{z}) = N |\mathcal{V}|,
$$
where $|\mathcal{V}|$ denotes the cardinality of the neighbourhood structure $\mathcal{V}$. For the Potts MRF this complement is given by
\begin{eqnarray} \label{CHamiltonian}
    H^*(\boldsymbol{z}) \triangleq \sum_{n=1}^N\sum_{n'\in \mathcal{V}(n)} \delta(z_{n} \neq z_{n'}).
\end{eqnarray}
Replacing $H(\boldsymbol{z}) =  N |\mathcal{V}| - H^*(\boldsymbol{z})$ in \eqref{posterior3} we obtain 
\begin{eqnarray} \label{posterior4}
\begin{split}
f\left(\bx,\bz,\bmu,\beta |\boldsymbol{y}\right) &\propto
								\left(\prod_{k=1}^K \,\,\prod_{n \in \mathcal{S}_k} p_{\mathcal{N}}(y_n|x_n,\sigma^2)\delta(x_n - \mu_k)\right) f(\bmu) \left[H^*(\boldsymbol{z}) + (\gamma - N |\mathcal{V}|)\right]^{-(\alpha + N)}.
\end{split}
\end{eqnarray}

 
Furthermore, noting that $H^*(\boldsymbol{z})$ only measures if neighbour labels are identical or not, regardless of their values, it is easy to check that the posterior \eqref{posterior3} remains unchanged if we substitute $H^*(\boldsymbol{z})$ with $H^*(\boldsymbol{x})$ 
\begin{eqnarray}
\begin{split}
f\left(\bx,\bz,\bmu,\beta |\boldsymbol{y}\right) 	&\propto
								f(\bmu) \left[H^*(\boldsymbol{x}) +  (\gamma - N |\mathcal{V}|)\right]^{-(\alpha + N)} \prod_{k=1}^K \,\,\prod_{n \in \mathcal{S}_k} p_{\mathcal{N}}(y_n|x_n,\sigma^2)\delta(x_n - \mu_k).
\end{split}
\end{eqnarray}
Finally, we make the observation that for 1st order neighbourhoods (see Fig. \ref{fig:neighborhood}) we have $H^*(\boldsymbol{\bx}) = 2||\nabla \bx||_0$, where $||\nabla \bx||_0 \,= ||\nabla_h \bx||_0 + ||\nabla_v \bx||_0$ denotes the $\ell_0$ norm of the horizontal and vertical components of the 1st order discrete gradient of $\bx$, and therefore
\begin{eqnarray} \label{posterior5}
\begin{split}
f\left(\bx,\bz,\bmu,\beta |\boldsymbol{y} \right) 	&\propto
								f(\bmu) \left[||\nabla \bx||_0 + (\gamma - N |\mathcal{V}|)/2 \right]^{-(\alpha + N)} \prod_{k=1}^K \,\,\prod_{n \in \mathcal{S}_k} p_{\mathcal{N}}(y_n|x_n,\sigma^2)\delta(x_n - \mu_k).
\end{split}
\end{eqnarray}

The graphical structure of this equivalent hierarchical Bayesian model is summarised in Fig. \ref{fig:DAG3} below. Notice that in this model $\bx$ separates $\by$ and $\sigma^2$ from the other model parameters, that the regularisation parameter $\beta$ has been marginalised, that the MRF is now enforcing spatial smoothness on $\bx$ not $\bz$, and that the elements of $\bz$ are prior independent.
\begin{figure}[h!]
\begin{minipage}[a1]{.48\linewidth}
\centerline{ \xymatrix{
*+[F]{\alpha,\gamma}\ar@{->}[d] & *+[F]{\rho^2}\ar@{->}[d] & \\
*+<0.04in>+[F.] +{\beta}\ar@{->}[dr] & \bmu\ar[d] & \bz \ar[dl]\\
\sigma^2 & \bx\ar@{->}[d] & \\
& \by\ar@{<-}[ul] \\
}
}
\end{minipage}
\begin{minipage}[a1]{.48\linewidth}
\xymatrix@!0{
\bz: & & z_n \ar@{->}'[d][ddd]
& & z_{n+1} \ar@{->}[ddd]]& & 
\\
& z_{n^\prime}  \ar@{->}[ddd]
& & z_{n^\prime+1}  \ar@{->}[ddd]& 
\\
& & & & & 
\\
\bx: & & x_n \ar@{.}[l] \ar@{.}[ur]\ar@{-}[rr]\ar@{->}'[d][ddd]
& & x_{n+1} \ar@{->}[ddd]\ar@{.}[r] \ar@{.}[ur]& & 
\\
& x_{n^\prime}  \ar@{.}[dl]\ar@{.}[l] \ar@{-}[ur]\ar@{-}[rr]\ar@{->}[ddd]
& & x_{n^\prime+1}  \ar@{-}[ur]\ar@{->}[ddd]\ar@{.}[r]\ar@{.}[dl]& 
\\
& & & & & 
\\
\by: & &  y_n 
& & y_n
\\
& y_{n^\prime} 
& & y_{n^\prime+1}
}
\end{minipage}
 \caption{[Left:] Directed acyclic graph of the proposed Bayesian model, augmented by the auxiliary variable $\bx$ decoupling $\bmu$ and $\bz$ from $\by$, and with marginalisation of the regularisation parameter $\beta$ (parameters with fixed values are represented using solid black boxes, marginalised variables appear in dashed boxes). [Right] Local representation of three layers of the model for $4$ neighbouring pixels.} \label{fig:DAG3}
\end{figure}
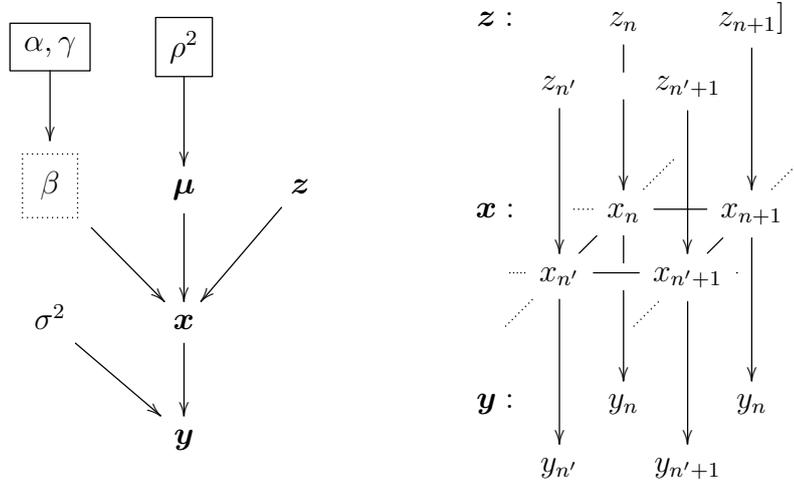


We are now ready to conduct a small-variance asymptotics analysis on \eqref{posterior5} and derive the asymptotic MAP estimator of $\bx,\bz,\bmu$, which is defined for our model as \cite{Broderick2013}
$$
\argmin_{\bx,\bz,\bmu}\, \lim_{\sigma^2 \rightarrow 0} -\sigma^2\log f\left(\bx,\bz,\bmu |\boldsymbol{y} \right).
$$
First, we use the fact that $\delta(s) = \lim_{\tau^2 \rightarrow 0} p_{\mathcal{N}}(s|0,\tau^2)$ to express \eqref{posterior5} as follows
\begin{eqnarray} \label{posterior6}
\begin{split}
&f\left(\bx,\bz,\bmu |\boldsymbol{y},\beta\right) \\
	&\quad\propto \lim_{\tau^2 \rightarrow 0}
								\left(\prod_{k=1}^K \,\,\prod_{n \in \mathcal{S}_k} p_{\mathcal{N}}(y_n|x_n,\sigma^2) p_{\mathcal{N}}(x_n | \mu_k,\tau^2)\right)\\
								&\quad\quad\times f(\bmu) \left[||\nabla \bx||_0 + (\gamma - N |\mathcal{V}|)/2 \right]^{\alpha + N},\\
								&\quad\propto \lim_{\tau^2 \rightarrow 0}
								\left(\prod_{k=1}^K \,\,\prod_{n \in \mathcal{S}_k} \exp\left({-\frac{(x_n-y_n)^2}{2\sigma^2}-\frac{(x_n - \mu_k)^2}{2\tau^2}}\right)\right)\\
								&\quad\quad\times f(\bmu) \left[||\nabla \bx||_0 + (\gamma - N |\mathcal{V}|)/2 \right]^{-(\alpha + N)}.
								\end{split}
\end{eqnarray}
Then, in a manner akin to Broderick et al. \cite{Broderick2013}, we allow the model's hyper parameters to scale with $\sigma^2$ in order to preserve the balance between the prior and the likelihood and avoid a trivial limit. More precisely, we set $\alpha = N /\sigma^2$ and assume that $\sigma^2$ vanishes at the same speed as $\tau^2$. Then, the limit of $-\sigma^2 \log f\left(\bx,\bz,\bmu |
\boldsymbol{y}\right)$ as $\sigma^2 \rightarrow 0$ is given by
\begin{eqnarray}
\begin{split}
\lim_{\sigma^2 \rightarrow 0} -\sigma^2\log f\left(\bx,\bz,\bmu |\boldsymbol{y}\right) = &\sum_{k=1}^K \,\,\sum_{n \in \mathcal{S}_k} \frac{1}{2}(x_n-y_n)^2 +\frac{1}{2}(x_n - \mu_k)^2\\
															     & + N\log(||\nabla \bx||_0 + (\gamma - N |\mathcal{V}|)/2),
\end{split}
\end{eqnarray}
and the MAP asymptotic estimators of $\bx,\bz,\bmu$ by
\begin{eqnarray} \label{SVA}
\begin{split}
\argmin_{\bx,\bz,\bmu}\, &\sum_{k=1}^K \,\,\sum_{n \in \mathcal{S}_k} \frac{1}{2}(x_n-y_n)^2 +\frac{1}{2}(x_n - \mu_k)^2 + N\log(||\nabla \bx||_0 + 1),
\end{split}
\end{eqnarray}
where we have set $\gamma = 2 + N |\mathcal{V}|$ such that the penalty $\log \left[||\nabla \bx||_0 + (\gamma - N |\mathcal{V}|)/2 \right] \geq 0$.

\subsection{Convex relaxation and optimisation}
Computing the estimator \eqref{SVA} is still NP-hard due to $\log(||\nabla \bx||_0 +1)$. To address this difficulty we use a convex relaxation of $||\nabla \bx||_0$ and exploit the concavity of the logarithmic function. Precisely, we replace $||\nabla \bx||_0$ by the convex approximation $\textrm{TV}(\bx) = ||\nabla \bx||_{1-2}$, (i.e.,  the isotropic total-variation pseudo-norm of $\bx$ \cite{Rudin:1992}), and obtain the following optimisation problem
\begin{eqnarray} \label{SVA2}
\begin{split}
\argmin_{\bx,\bz,\bmu}\, &\sum_{k=1}^K \,\,\sum_{n \in \mathcal{S}_k} \frac{1}{2}(x_n-y_n)^2 +\frac{1}{2}(x_n - \mu_k)^2 + N\log(TV(\bx) + 1),
\end{split}
\end{eqnarray}
which can be very efficiently computed by iterative minimisation w.r.t. $\bx$, $\bz$ and $\bmu$. The minimisation of \eqref{SVA2} w.r.t. $\bz$ (with $\bx$ and $\bmu$ fixed) is a trivial separable integer problem that can be formulated as $N$ independent (pixel-wise) minimisation problems over $1,\ldots,K$ (these unidimensional integer problems can be solved by simply checking the value $z_n = 1,\ldots,K$ that minimises \eqref{SVA2} for each pixel $n = 1,\ldots,N$). Similarly, the minimisation with respect to $\bmu$ is a trivial quadratic least squares fitting problem with analytic solution (i.e., by setting $\mu_k = \frac{1}{|S_k|}\sum_{n \in \mathcal{S}_k} x_n$ for each $k = 1,\ldots, K$, where $|S_k|$ denotes the cardinality of $S_k$). Also note that iteratively minimising \eqref{SVA2} with respect to $\bz$ and $\bmu$, with fixed $\bx$, is equivalent to solving a least squares clustering problem with the popular K-means algorithm \cite{macqueen:1967}. Moreover, the minimisation of \eqref{SVA2} w.r.t. $\bx$ (with $\bz$ and $\bmu$ fixed) is achieved by solving the non-convex optimisation problem
\begin{eqnarray}\label{non-convex}
\begin{split}
\argmin_{\bx}\, \sum_{k=1}^K \,\,\sum_{n \in \mathcal{S}_k} &\frac{1}{2}(x_n-y_n)^2 +\frac{1}{2}(x_n - \mu_k)^2 + N \log\left[TV(\bx) \ + 1\right],
\end{split}
\end{eqnarray} 
which was studied in detail in \cite{oliveira:2009}. Essentially, given some initial condition $\bv^{(0)} \in \mathbb{R}^N$, \eqref{non-convex} can be efficiently minimised by majorisation-minimisation (MM) by iteratively solving the following sequence of trivial convex problems,
\begin{eqnarray}\label{MM}
\begin{split}
\bv^{(\ell+1)} = \argmin_{\bx}\, &\sum_{k=1}^K \,\,\sum_{n \in \mathcal{S}_k} \frac{1}{2}(x_n-y_n)^2 +\frac{1}{2}(x_n - \mu_k)^2 + \lambda_\ell TV(\bx),\\ 
& \textrm{with } \lambda_\ell = \frac{N}{TV[\bv^{(\ell)}] +1},
\end{split}
\end{eqnarray} 
in which $\lambda_\ell$ plays the role of a regularisation parameter, and where we have used the majorant \cite{oliveira:2009}
\begin{eqnarray}
\begin{split}
q(\bx | \bv^{(\ell)}) &= \frac{\left(TV(\bx) - TV(\bv^{(\ell)})\right)}{(TV(\bv^{(\ell)}) + 1)} +  \log\left(TV(\bx) + 1 \right)\\
		     &\geq \log\left(TV(\bv^{(\ell)}) + 1 \right).
\end{split}
\end{eqnarray}
Notice that each step of \eqref{MM} is equivalent to a trivial convex total-variation denoising problem that can be very efficiently solved, even in high-dimensional scenarios, by using modern convex optimisation techniques (in this paper we used a parallel implementation of Chambolle's algorithm \cite{chambolle:2004}).

The proposed unsupervised segmentation algorithm based on \eqref{SVA2} is summarised in Algo. \ref{Algo2} below. We note at this point that because the overall minimisation problem is not convex the solution obtained by iterative minimisation of \eqref{SVA2} might depend on the initial values of $\bx,\bz,\bmu$. In all our experiments we have used the initialisation $\bx^{(0)} = 2\by$, $\bz = [1,\ldots,1]^T$, $\bmu = [0,\ldots,0]^T$ that produced good estimation results. 
\begin{algorithm}{}
     \caption{Unsupervised Bayesian segmentation algorithm}
     \begin{algorithmic}[1]
	\STATE \underline{Input:} Image $\by$, number of maximum outer iterations $T$ and inner iterations $L$, tolerance level $\epsilon$.
        \STATE Initialise $\bx^{(0)} = 2\by$, $\bz = [1,\ldots,1]^T$, $\bmu = [0,\ldots,0]^T$.
        \FOR{$t=1:T$}
        \STATE Set $\bv^{(0)} = \bx^{(t-1)}$.
        \FOR{$\ell=0:L$}
			\STATE Set $\lambda_\ell= N/\{TV[\bv^{(\ell)}] +1\}$.
			\STATE Compute $\bv^{(\ell+1)}$ using \eqref{MM}, with fixed $\bz = \bz^{(t-1)}$ and $\bmu = \bmu^{(t-1)}$, using Chambolle's algorithm \cite{chambolle:2004}.
			\IF{$(N/\{TV[\bv^{(\ell+1)}] +1\}-\lambda) \geq \epsilon \lambda$}
				\STATE Set $\ell = \ell +1$.
			\ELSE
				\STATE Exit to line 14.
			\ENDIF
       	\ENDFOR
	\STATE Set $\bx^{(t)} = \bv^{(L)}$.
	\STATE Compute $\bz^{(t)}$ and $\bmu^{(t)}$ by least-squares clustering of $\bx^{(t)}$ using the K-means algorithm \cite{macqueen:1967}.
	\IF {$\bz^{(t)} \neq \bz^{(t-1)}$}
		\STATE  Set $t = t+1$.
	\ELSE
		\STATE Exit to line 22.
	\ENDIF
       	\ENDFOR
       	\STATE \underline{Output:} Segmentation $\bz^{(t)}$, $\bmu^{(t)}$, $\lambda = N/(TV[\bx^{(t)}] +1)$.
\end{algorithmic}     \label{Algo2}
\end{algorithm}

%
\section{Experimental Results and Observations}\label{sec:experiments}
In this section we demonstrate empirically the proposed Bayesian image segmentation methodology with a series of experiments and comparisons with state-of-the-art algorithms. To asses the accuracy of our method we compare the results with the estimations produced by the Markov chain Monte Carlo algorithm \cite{Pereyra_TIP_2013}, which estimates the marginal posterior of the segmentation labels $f\left(\bz|\boldsymbol{y}\right)$ with very high accuracy. We also report comparisons with four supervised fast image segmentation techniques that we haven chosen to represent different efficient algorithmic approaches to image segmentation (e.g. MRF energy minimisation solved by graph-cut, active contour solved by Riemannian gradient descent, and two convex models solved by convex optimisation). The specific methods used in the comparison are as follows:
\begin{itemize}
\item{ The two-stage smoothing-followed-by-thresholding algorithm (TSA) \cite{Raymond2013}, which is closely related to a semi-supervised instance of Algo. \ref{Algo2} with a single iteration (TV-denoising followed by K-means), and with a fixed regularisation parameter $\lambda$ specified by the practitioner.}
\item{Hidden Potts MRF segmentation \eqref{posterior} with fixed $\beta$, solved by graph max-flow/min-cut approximation \cite{Boykov2004}.} 
\item{Chan-Vese active contour by natural gradient descent \cite{PereyraSTSP2013} (to our knowledge this method is currently the fastest approach for solving active contour models).}
\item{The fast global minimisation algorithm (FGMA) \cite{Bresson2007} for active contour models. In a similar fashion to our method, this algorithm also involves a model with a TV convex relaxation that is solved by convex optimisation.}
\end{itemize}
We emphasise that, unlike the proposed method, all these efficient approaches are supervised, i.e., they require the specification of a regularisation parameters. In the experiments reported hereafter we have tuned and adjusted the parameters of each algorithm to each image by use of visual cross-validation to ensure we produce the best results for each method on each image.

\begin{figure}[htb]
\begin{minipage}[a2]{.99\linewidth}
  \centering
  \centerline{\includegraphics[width=10cm]{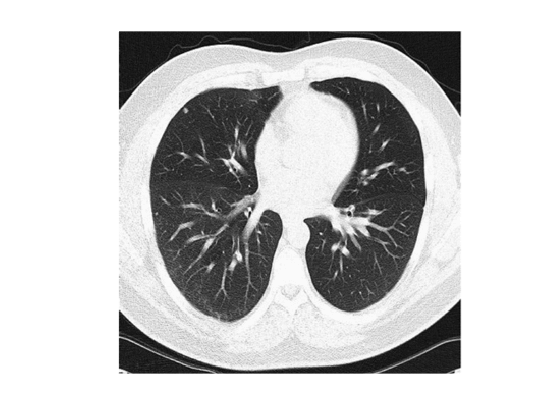}}
    \centerline{(a)  {\texttt{Lung}}}\medskip
\end{minipage}
\begin{minipage}[a1]{.99\linewidth}
  \centering
  \centerline{\includegraphics[width=10cm]{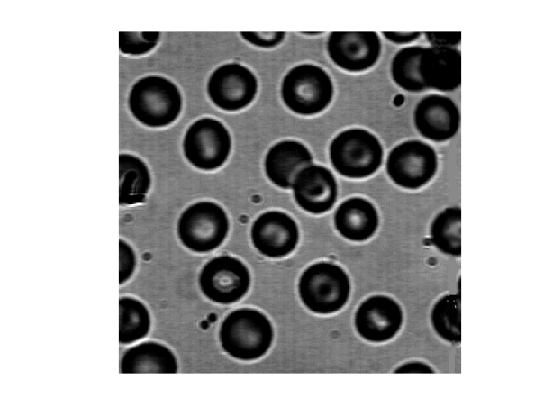}}
  \centerline{(b)  \texttt{Bacteria}}\medskip
\end{minipage}
\hfill
\begin{minipage}[a1]{.99\linewidth}
  \centering
  \centerline{\includegraphics[width=10cm]{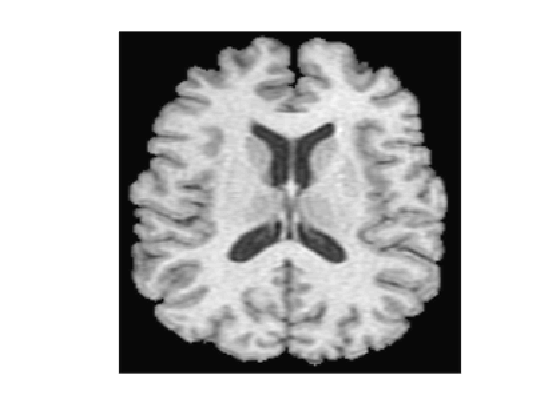}}
  \centerline{(c)  {\texttt{Brain}}}\medskip
\end{minipage}
\caption{\small{The \texttt{Lungs} ($336 \times 336$ pixels), \texttt{Bacteria} ($380 \times 380$ pixels), and \texttt{Brain} ($256 \times 256$ pixels) images used in the experiments.}}
\label{Fig_1a}
\end{figure}

To guarantee that the comparisons are fair we have applied the six algorithms considered in this paper to three images with very different compositions: the \texttt{Lungs} and \texttt{Bacteria} images from the supplementary material of \cite{Bresson2007}, and one slice of a 3D in-vivo MRI image of a human brain composed by biological tissues (white matter and grey matter) with complex shapes and textures, making the segmentation problem challenging. The three test images are depicted in Figure \ref{Fig_1a}. These images have been selected as they are composed of different types and numbers of objects; objects which have different shapes, (regular and irregular); and a range of potential segmentation solutions. All experiments have been conducted using a MATLAB implementation of Algo. \ref{Algo2} with parameters $T = 50$, $L = 25$, $\epsilon = 10^{-3}$, and computed on an Intel i7 quad-core workstation running MATLAB 2014a.  With regards to the algorithms used for comparison, when possible we have used MATLAB codes made available by the respective authors. It should be noted that these are mainly MATLAB scripts, however the graph-cut method is written in C++, ( the \cite{Bagon2006} implementation was used here), so it has a slight advantage in terms of computational performance.

We emphasise at this point that we do not seek to explicitly compare the accuracy of the methods because: 1) there is no objective ground truth; 2) the "correct" segmentation is often both subjective and application-specific; and 3) the segmentations can often be marginally improved by fine tuning the regularisation parameters. What our experiments seek to demonstrate is that our method performs similarly to the most efficient deterministic approaches of the state-of-the-art, both in terms of segmentation results and computing speed, with the fundamental advantage that it does not require specification of the value of regularisation parameters (i.e., it is fully unsupervised).

\begin{figure}[htb]

\begin{minipage}[a2]{.49\linewidth}
  \centering
  \centerline{\includegraphics[width=9cm]{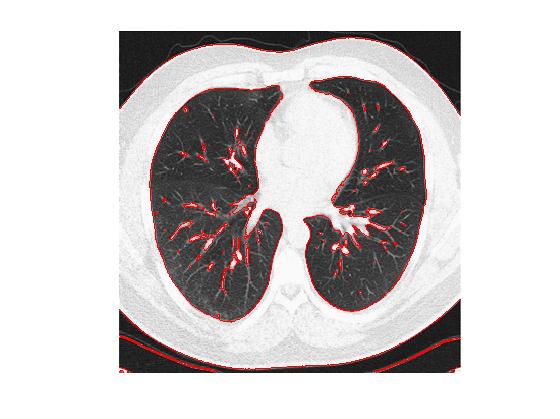}}
    \centerline{(a)  \small{Proposed}}\medskip
\end{minipage}
\begin{minipage}[a1]{.49\linewidth}
  \centering
  \centerline{\includegraphics[width=9cm]{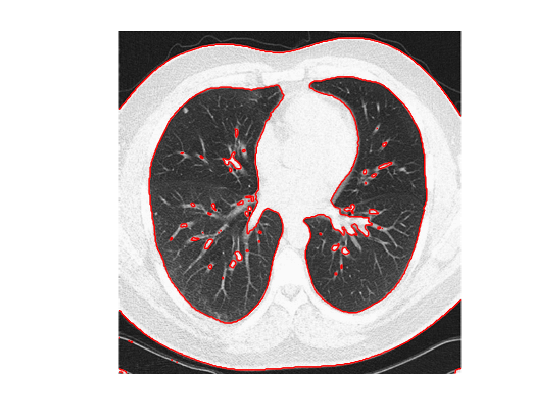}}
  \centerline{(b)  \small{MCMC \cite{Pereyra_TIP_2013}}}\medskip
\end{minipage}
\hfill
\begin{minipage}[a1]{.49\linewidth}
  \centering
  \centerline{\includegraphics[width=9cm]{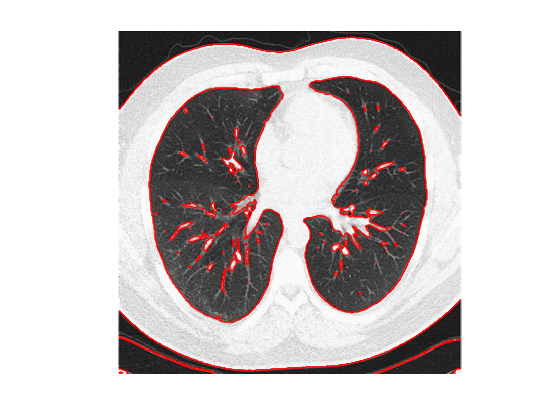}}
  \centerline{(c)  \small{TSA \cite{Raymond2013}}}\medskip
\end{minipage}
\begin{minipage}[a1]{.49\linewidth}
  \centering
  \centerline{\includegraphics[width=9cm]{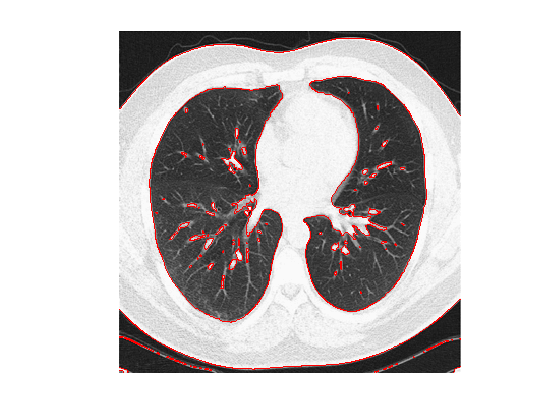}}
   \centerline{(f)  \small{Graph-Cut \cite{Boykov2004}}}\medskip
\end{minipage}
\hfill
\begin{minipage}[a1]{.49\linewidth}
  \centering
  \centerline{\includegraphics[width=9cm]{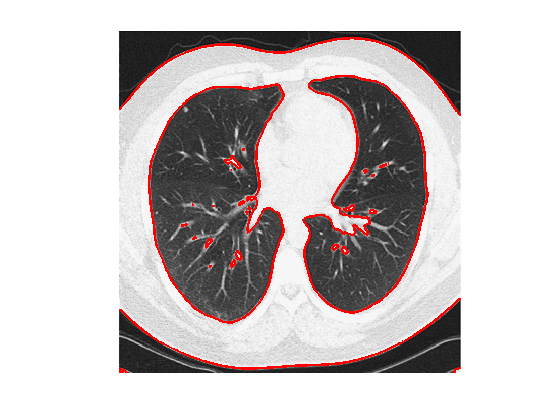}}
  \centerline{(e)  \small{Natural grad. \cite{PereyraSTSP2013}}}\medskip
\end{minipage}
\begin{minipage}[a2]{.49\linewidth}
  \centering
  \centerline{\includegraphics[width=9cm]{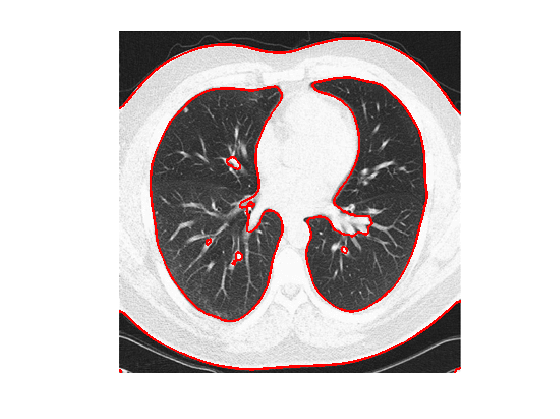}}
  \centerline{(f)  \small{FGMA \cite{Bresson2007}}}\medskip
\end{minipage}
\caption{\small{Comparison with the state-of-the-art methods \cite{PereyraSTSP2013}, \cite{Boykov2004}, \cite{Bresson2007}, and \cite{Raymond2013} using the lung image ($336 \times 336$ pixels) from the supplementary material of \cite{Bresson2007}.}}
\label{Fig_1b}
\end{figure}

\begin{figure}[htb]
\begin{minipage}[a2]{.49\linewidth}
  \centering
  \centerline{\includegraphics[width=9cm]{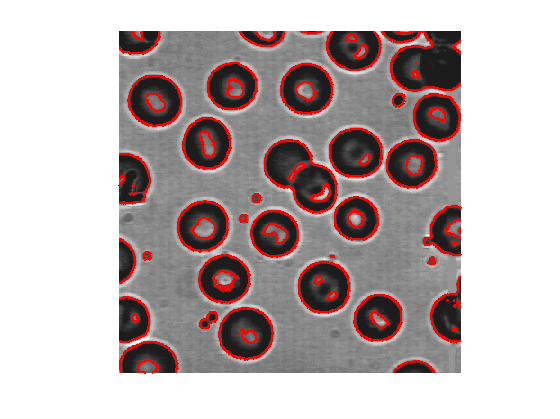}}
  \centerline{(a)  \small{Proposed}}\medskip
\end{minipage}
\begin{minipage}[a1]{.49\linewidth}
  \centering
  \centerline{\includegraphics[width=9cm]{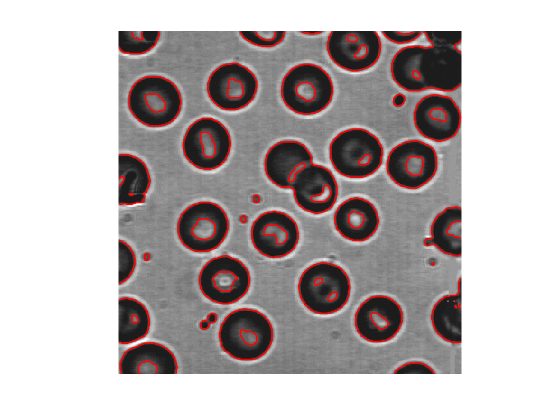}}
  \centerline{(b)  \small{MCMC \cite{Pereyra_TIP_2013}}}\medskip
\end{minipage}
\hfill
\begin{minipage}[a1]{.49\linewidth}
  \centering
  \centerline{\includegraphics[width=9cm]{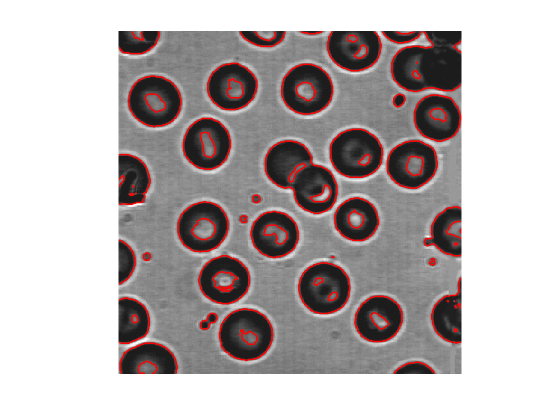}}
    \centerline{(c)  \small{TSA \cite{Raymond2013}}}\medskip
\end{minipage}
\begin{minipage}[a1]{.49\linewidth}
  \centering
  \centerline{\includegraphics[width=9cm]{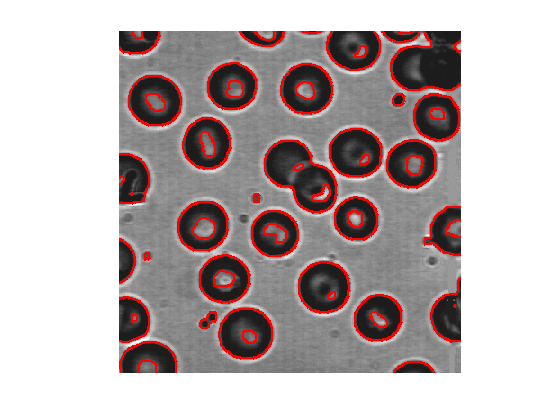}}
  \centerline{(f)  \small{Graph-Cut \cite{Boykov2004}}}\medskip
\end{minipage}
\begin{minipage}[a1]{.49\linewidth}
  \centering
  \centerline{\includegraphics[width=9cm]{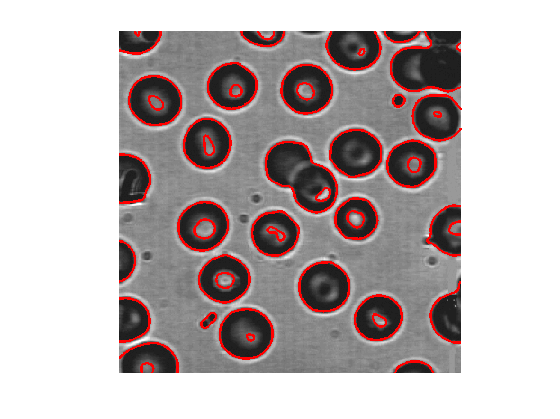}}
   \centerline{(e)  \small{Natural gradient \cite{PereyraSTSP2013} }}\medskip
\end{minipage}
\hfill
\begin{minipage}[a2]{.49\linewidth}
  \centering
  \centerline{\includegraphics[width=9cm]{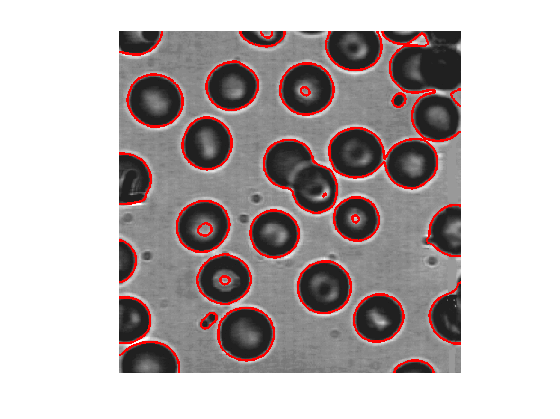}}
  \centerline{(f)  \small{FGMA \cite{Bresson2007}}}\medskip
\end{minipage}
\caption{\small{Comparison of the supervised and unsupervised methods with the state of the algorithm \cite{PereyraSTSP2013}, \cite{Boykov2004}, \cite{Bresson2007} and \cite{Raymond2013} using the bacteria image ($380 \times 380$ pixels) from the supplementary material of \cite{Bresson2007}.}}
\label{Fig_1c}
\end{figure}

\begin{figure}[htb]
  \begin{minipage}[a1]{.49\linewidth}
  \centerline{\includegraphics[width=9cm]{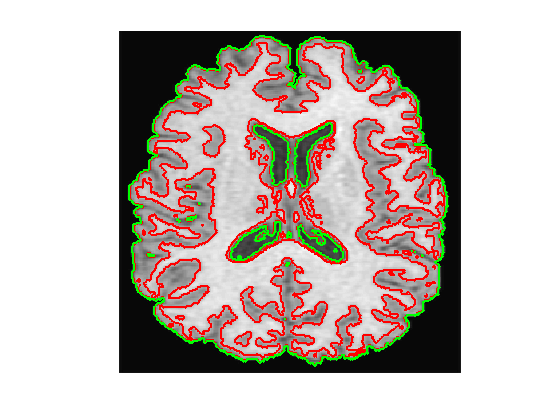}}
  \centerline{(a)  \small{Proposed}}\medskip
  \end{minipage}
  \hfill
  \begin{minipage}[a1]{.49\linewidth}
      \centerline{\includegraphics[width=9cm]{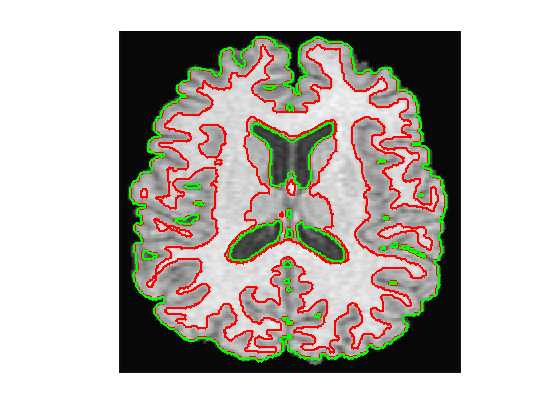}}
  \centerline{(b)  \small{MCMC \cite{Pereyra_TIP_2013}}}\medskip
  \end{minipage}
\begin{minipage}[a1]{.49\linewidth}
  \centerline{\includegraphics[width=9cm]{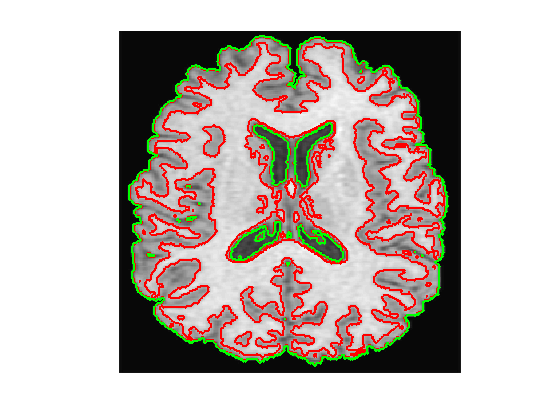}}
  \centerline{(c)  \small{TSA \cite{Raymond2013}}}\medskip
  \end{minipage}
  \hfill
  \begin{minipage}[a1]{.49\linewidth}
    \centerline{\includegraphics[width=9cm]{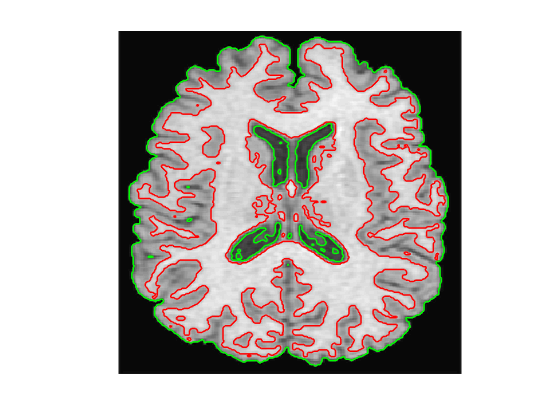}}
  \centerline{(d)  \small{Graph-Cut \cite{Boykov2004}}}\medskip
    \end{minipage}

\caption{\small{Segmentation of a brain MRI image ($256 \times 256$ pixels).}}
\label{Fig_2}
\end{figure}

Figures \ref{Fig_1b}, \ref{Fig_1c}, and \ref{Fig_2} respectively show the segmentation results obtained for the \texttt{Lungs}, \texttt{Bacteria} and \texttt{Brain} test images with each method. The segmentations of the \texttt{Lungs} and \texttt{Bacteria} images have been computed using $K=2$ classes to enable comparison with the natural gradient method \cite{PereyraSTSP2013} and FGMA \cite{Bresson2007} (these methods are based on an active contour model that only supports binary segmentations), whereas the \texttt{Brain} image has been computed using $K = 3$ classes to produce a clear segmentation of the grey matter and the white matter. The computing times associated with these experiments are reported in Table \ref{Tab3b}. Observe that all six methods produced similar segmentation results that are in good visual agreement with each other. In particular, we observe that the proposed method successfully determined the appropriate level of regularisation for each image and produced segmentations that are very similar to the results obtained with the supervised methods graph-cut \cite{Boykov2004} and TSA \cite{Raymond2013}, and with the unsupervised MCMC algorithm \cite{PereyraSTSP2013} that in a sense represents a benchmark for these approximate inference methods. Moreover, Table \ref{Tab3b} shows that the proposed method was only $2$ or $3$ times slower than state-of-the-art supervised approaches, which is an excellent performance for a fully unsupervised method. This additional computing time is mainly due to the additional computations related to the non-convex program \eqref{non-convex}; however, we emphasise that this algorithm has the property of adapting automatically the level of regularisation to the image, and that the computing times reported in Table \ref{Tab3b} do not take into account the time involved in running the supervised algorithms repeatedly to adjust their regularisation parameters.

\begin{table}[htb]
\renewcommand{\arraystretch}{1.3}
\caption{Computing times (seconds) for the \texttt{Lungs}, \texttt{Bacteria} and \texttt{Brain} images displayed in Figs. \ref{Fig_1b}, Figs. \ref{Fig_1c} and Figs. \ref{Fig_2}.}
\label{Tab3b}
\centering
\begin{tabular}{|c||c|c|c|}
  \cline{2-4}
  \multicolumn{1}{c|}{}                        & \texttt{Bacteria} & \texttt{Bacteria} & \texttt{Brain}\\
  \hline
  \hline
  Proposed    							& $0.65$ & $0.80$ & $0.23$\\
  \hline
  TSA \cite{Raymond2013}   				& $0.20$ & $0.21$  & $0.17$\\
  \hline
  Graph-Cut \cite{Boykov2004}    			& $0.30$ & $0.30$ & $0.21$\\
  \hline
  Natural gradient \cite{PereyraSTSP2013}   	& $0.20$ & $0.18$ & n/a\\
  \hline
  FGMA \cite{Bresson2007}    		            	& $0.32$ & $0.47$ & n/a\\
  \hline
  MCMC \cite{PereyraSTSP2013}			& $900$ & $1\,150$ & $533$\\ 
  \hline
  \end{tabular}
\end{table}

\section{Conclusions}
We have presented a new fully unsupervised approach for computationally efficient image segmentation. The approach is based on a new approximate Bayesian estimator for hidden Potts-Markov random fields with unknown regularisation parameter $\beta$. The estimator is based on a small-variance-asymptotic analysis of an augmented Bayesian model and a convex relaxation combined with majorisation-minimisation technique. This estimator can be very efficiently computed by using an alternating direction scheme based on a convex total-variation denoising step and a least-squares (K-means) clustering step, both of which can be computed straightforwardly, even in large $2$D and $3$D scenarios, and with parallel computing techniques. Experimental results on real images, as well as extensive comparisons with state-of-the-art algorithms showed that the resulting new image segmentation methodology performs similarly in terms of segmentation results and of computing times as the most efficient supervised image segmentation methods, with the important additional advantage of self-adjusting regularisation parameters. A detailed analysis of the theoretical properties of small-variance-asymptotic estimators in general, and in particular of the methods described in this paper, is currently under investigation. Potential future research topics include the extension of these methods to non-Gaussian statistical models from the exponential family and their application to ultrasound and PET image segmentation, extensions to models with unknown number of classes $K$, and comparisons with other Bayesian segmentation methods based on alternative hidden MRF models that can also be solved by convex optimisation, such as \cite{bioucas:2014}.

\footnotesize
\bibliographystyle{ieeetran}
\bibliography{strings,strings2}
\end{document}